\documentclass[utf8]{article} 
\usepackage{url,hyperref,lineno,microtype,booktabs}
\usepackage[onehalfspacing]{setspace}
\usepackage{longtable}
\usepackage{fullpage}
\usepackage{authblk}
\usepackage{graphicx}

\begin{document}

\title{Optic Disc Segmentation using Disk-Centered Patch Augmentation}

\date{}



\author{Saeid Motevali}
\author{Aashis Khanal}
\author{Rolando Estrada}
\affil{\texttt{\{smotevalialamoti1, akhanal1\}@student.gsu.edu, restrada1@gsu.edu}}

\affil{Department of Computer Science, Georgia State University, GA, USA}

\maketitle

\begin{abstract}
The optic disc is a crucial diagnostic feature in the eye since changes to its physiognomy is correlated with the severity of various ocular and cardiovascular diseases. While identifying the bulk of the optic disc in a color fundus image is straightforward, accurately segmenting its boundary at the pixel level is very challenging. In this work, we propose disc-centered patch augmentation (DCPA)|a simple, yet novel training scheme for deep neural networks|to address this problem. DCPA achieves state-of-the-art results on full-size images even when using small neural networks, specifically a U-Net with only 7 million parameters as opposed to the original 31 million. In DCPA, we restrict the training data to patches that fully contain the optic nerve. In addition, we also train the network using dynamic cost functions to increase its robustness. We tested DCPA-trained networks on five retinal datasets: DRISTI, DRIONS-DB, DRIVE, AV-WIDE, and CHASE-DB. The first two had available optic disc ground truth, and we manually estimated the ground truth for the latter three. Our approach achieved state-of-the-art F1 and IOU results on four datasets (95\% F1, 91\% IOU on DRISTI; 92\% F1, 84\% IOU on DRIVE; 83\% F1, 71\% IOU on AV-WIDE; 83\% F1, 71\% IOU on CHASEDB) and competitive results on the fifth (95\% F1, 91\% IOU on DRIONS-DB), confirming its generality. Our open-source code and ground-truth annotations are available at: \url{https://github.com/saeidmotevali/fundusdisk}.




\end{abstract}

\section{Introduction}
\label{sec:introduction}
The optic disc|the region where ganglion cell axons and blood vessels exit the retina|is a crucial diagnostic structure in the eye since changes to its physiognomy are correlated with the severity of various diseases, including glaucoma \cite{VARMA1992215}, idiopathic intracranial hypertension (IIH) \cite{doi:10.1111/j.1526-4610.2008.01324.x}, coronary heart disease \cite{10.1093/eurheartj/ehm221}, and atherosclerosis \cite{10.1167/iovs.03-1390}. For example, the cup-to-disc ratio (CDR), which measures the ratio of the disc to its central depression (cup), is widely used to diagnose glaucoma \cite{4408713, Walter2001SegmentationOC, Zhu2010}. Glaucomatous eyes show more pronounced cupping due to an increase in intracranial pressure, which causes flowing nutrients to pass through damaged axons and induce swelling of the optic disc \cite{1687814016649298}. Optic disc swelling can also be a sign of a brain tumor or a minor nerve stroke.

Many methods have been proposed for automatically segmenting the optic disc in color fundus images (see Sec.~\ref{sec:relatedWork} for a brief overview). In general, the bulk of the optic disc is easy to detect since it is brighter and yellower than the rest of the retina \cite{Zhu2010}. However, automatically determining the disc's \textit{boundary} at the pixel level is challenging due to multiple factors. First, imaging artifacts such as motion or blur can obscure the edges of the disc. Second, retinal diseases can make this boundary more irregular, making it harder for an example-trained system (e.g., a deep neural network) to accurately trace it on non-healthy eyes. Also, existing datasets for this problem are small and relatively rare, even compared to other retinal segmentation problems. Finally, an additional challenge of optic disk segmentation is that the percentage of positive pixels is extremely low \cite{8451543,Man+16,6091537}. Specifically, the optic disc only constituted an average of 3.46\% of the image across the five datasets in our experiments (see Tbl.~\ref{tab:datasets} for details). 


To address these issues, we propose a novel scheme for training deep neural networks called \textit{disk center patch augmentation} (DCPA). Our approach is based on the U-net architecture \cite{Ronneberger2015UNetCN}, the current state-of-the-art deep architecture for medical segmentation problems. Specifically, in DCPA we restrict the training data to only include the region of the fundus image where the optic disc is located, improving the robustness of the trained model. In more detail, one can use a U-net-style network with a fixed input size for images of arbitrary size by splitting the input into \textit{patches}. These patches are fed to the neural network independently and then stitched together to form the final output. This patch-based approach is common in medical image segmentation. However, it is challenging to apply it to optic disc segmentation because the optic disk is significantly smaller than the fundus image as a whole. As such, the disc may be only present in a few of the patches. Patches without any positive pixels are problematic because (1) they do not contain any useful optic disc features, and (2) they can induce the network to become oversensitive to noise in the image. Thus, in DPCA we ensure that all the patches that we feed to the network during training contain the entire optic disc. We vary the position of the optic disc across patches to prevent location-based bias. Our implementation works with the original images (i.e., without resizing), thus it is compatible with downstream bio-markers for ocular disease diagnosis, such as the optic cup-to-disk ratio. 




In addition to DCPA, we allow make neural network training more robust by applying dynamic cost functions \cite{10.3389/fcomp.2020.00035} to this problem. Stochastic penalties allow a network to settle on an optimum that is more robust to ambiguous pixels (e.g., those at anatomical boundaries) than conventional training. In particular, a network trained with stochastic weights achieved state-of-the-art results in numerous retinal vessel segmentation datasets \cite{10.3389/fcomp.2020.00035}.


Training a deep neural network on medical images is very resource intensive due to their large size. In addition, it is not desirable to reduce the image size prior to processing because it might introduce unwanted pixel-level artifacts and/or loss of crucial information in the up/down-sampling process, both of which can lead to misdiagnoses. Fortunately, as we detail in Sec.~\ref{experiments_and_result}, our DCPA strategy allows us to achieve comparable results on OD segmentation using a much smaller network (at least 4.5 times smaller than a conventional U-net). We experimentally validated our approach across five retinal datasets|DRISTI, DRIVE, DRIONS-DB, AV-WIDE, and CHASE-DB|achieving state-of-the-art results on four of them and competitive results on the fifth. In addition to our comparison against the state of the art, we also carried out ablation studies on the different components of our system to better understand their impact on performance. Overall, we determined that centered patches consistently improved results, while the benefit of dynamic weights was more dataset specific.


The rest of this paper is organized as follows. In Sec.~\ref{sec:relatedWork}, we review prior work on automatic optic disc detection. We then detail our methodology in Sec.~\ref{sec:methodology} and present our experimental results in Sec.~\ref{experiments_and_result}, which we then discuss in Sec.~\ref{sec:discussion}. Finally, we conclude and note possible future directions in Sec.~\ref{sec:conclusions}.






\section{Related work}
\label{sec:relatedWork} 

On a color fundus image, the optic disc appears as a distinctive bright spot in the retina, located next to a dense, relatively dark cluster of vessels. Researchers have applied different techniques such as structural detection and convex hull \cite{7225107}, active contour \cite{doi:10.1117/12.708469} and edge filter \cite{6091537} to automatically segment this area. However, deep learning techniques are well known to learn low-to-high level features that generalize significantly better that other machine learning methods. \author{Mohan et al.} \cite{8451543} used CNN architecture, \author{Ronneberger et al.} \cite{Ronneberger2015UNetCN} used U-Net and \author{Maninis et al.} \cite{Man+16} applied deep retinal image understanding for OD segmentation. Below, we review some of the main techniques proposed for this problem.

\author{Roychowdhury et al.} in \cite{7225107} applied morphological reconstruction and applied a circular structure element on the green channel of the fundus image. Then the bright regions which is adjacent to major blood vessels detected as an optic disk. In the next step the binary classification applied to classify the bright region to OD and non-OD region. The area with maximum vessel-sum and solidity considered as a best candidate for OD. The other area within 1-disc diameter from the centroid of the best remaining OD candidate. After that the convex hull containing all the candidate OD regions is considered and the best-fit ellipse across the convex hull defines the segmentation OD boundary. 

\author{Kondermann et al.} in \cite{doi:10.1117/12.708469} used information of contrast and texture in image for detection of the optic disk. The model use statistical-based method to detect the optic disk contour in fundus retina images. After the initial guess of the contour, the method segments optic disk further by using active contour model (ACM) or snakes. A snake is a deformable boundary around an object for which the external energy of an image and the internal energy of the contour shape the final contour detection. There are a number of parameters that effects the final contour shape. Any variation in these parameters has a massive effect on the final contour detection. Also, such parameters needs to be tuned for each new dataset \cite{10.1007/978-3-540-24671-8_11}.

Peripapillary atrophy elimination was used in \cite{6091537} by \author{Cheng et al.} to detect the optic disk. The elimination applied edge filtering, constraint elliptical Hough transfer, and peripapillary atrophy detection. By applying this elimination, edges with higher probability of being non-disc structures especially peripapillary atrophy were excluded in order to achieve higher segmentation accuracy.

\author{Mohan et al.} in \cite{8451543} proposed Fine-Net which generates high-resolution optic disk segmentations of fundus images. Fine-Net is a CNN architecture with emphasis on localization accuracy. The framework generalized well even in test images with high variability. Meanwhile, \author{Maninis et al.} developed a deep retinal image understanding (DRIU) method in \cite{Man+16}, which provides both retinal vessel and optic disk segmentation. Their approach utilizes multiple deep convolutional neural networks (CNNs). In particular, it uses a base network architecture, as well as two sets of specialized layers to detect retinal vessels and segment the optic disk. 


\section{Methodology} 
\label{sec:methodology}

We now detail our novel scheme for training U-Net-style architectures \cite{Ronneberger2015UNetCN}. Our proposed approach consists of: (1) restricting the training data to relevant examples and (2) training the network using dynamic cost functions. Below, we first provide a quick overview of the U-Net architecture, then discuss each training component, in turn.



\subsection{U-net architecture}
The U-net architecture is an \textit{encoder-decoder} neural network with skip connections \cite{Ronneberger2015UNetCN}. In other words, it consists of a series of down-sampling layers followed by an equivalent number of up-sampling ones; the up-sampling layers match the size of the feature map from the down-sampling layer on the same level. This type of network can be applied to arbitrarily large images by scanning it across the full image. That is, one can partition the original image into \textit{patches} that fit the network's input size. In this case, the network's input is generally slightly larger than its output in order to utilize the context around the region of interest. For example, if the output is of size $w \times h$, the input will be $(w+p) \times (h+q)$, where the extra $p \times q$ region is a $w \times h$ pixel extension beyond the $p \times q$ center along each dimension. If the region extends beyond the edge of the image, standard practice is to mirror the pixels. As discussed further in Sec.~\ref{experiments_and_result}, we empirically determined the patch size for each dataset based on the average resolution of its images. 



As in \cite{10.3389/fcomp.2020.00035}, we use 3$ \times $3-pixel kernels except in the next-to-last layer, which has 1$ \times $1 kernels. The output layer is a \textit{softmax} layer with two probabilities: one for a pixel being a vessel and another for it being part of the background. A pixel is labeled as either \textit{vessel} or \textit{background} based on the higher of the two probabilities. Here, though, we use half the number of filters in each U-net layer compared to \cite{10.3389/fcomp.2020.00035}, for two reasons. First, unlike other segmentation problems (e.g., vessel segmentation), the optic disk only occupies a small portion of the image, thus requiring fewer features. Second, a smaller network can be trained faster and with less memory, allowing its use in less powerful hardware.

\subsection{Disk Centered Patch Augmentation}
\label{dcpa}
As noted above, the optic disc (OD) comprises a small percentage of a standard retinal image. Specifically, Tbl.~\ref{tab:datasets} shows that the average OD size can range from 9\% down to less than one percent across existing datasets. This small size makes it harder to train a neural network since only a small region around the OD is actually useful for segmenting it. Training a network on patches across the entire image (e.g., \cite{8451543}) forces the network to ignore a lot of noise in irrelevant patches in order to learn OD-specific features. To address this issue, we propose DCPA, a data selection method in which we select $r \times P$ random patches of size $w_i \times h_i$ from each image. Here, $r$ is a ratio between zero and one (we used $p=0.5$ for our experiments). For our final prediction, we picked the largest connected components from the network's output and filled any holes in this component via morphological operations. Importantly, we only use DCPA in the \textit{training phase} since we empirically found that processing the entire image during testing was more robust. Below, we describe how we process images during both training and testing.

\begin{figure}[ht!]
\begin{center}
\includegraphics[width=0.6\textwidth]{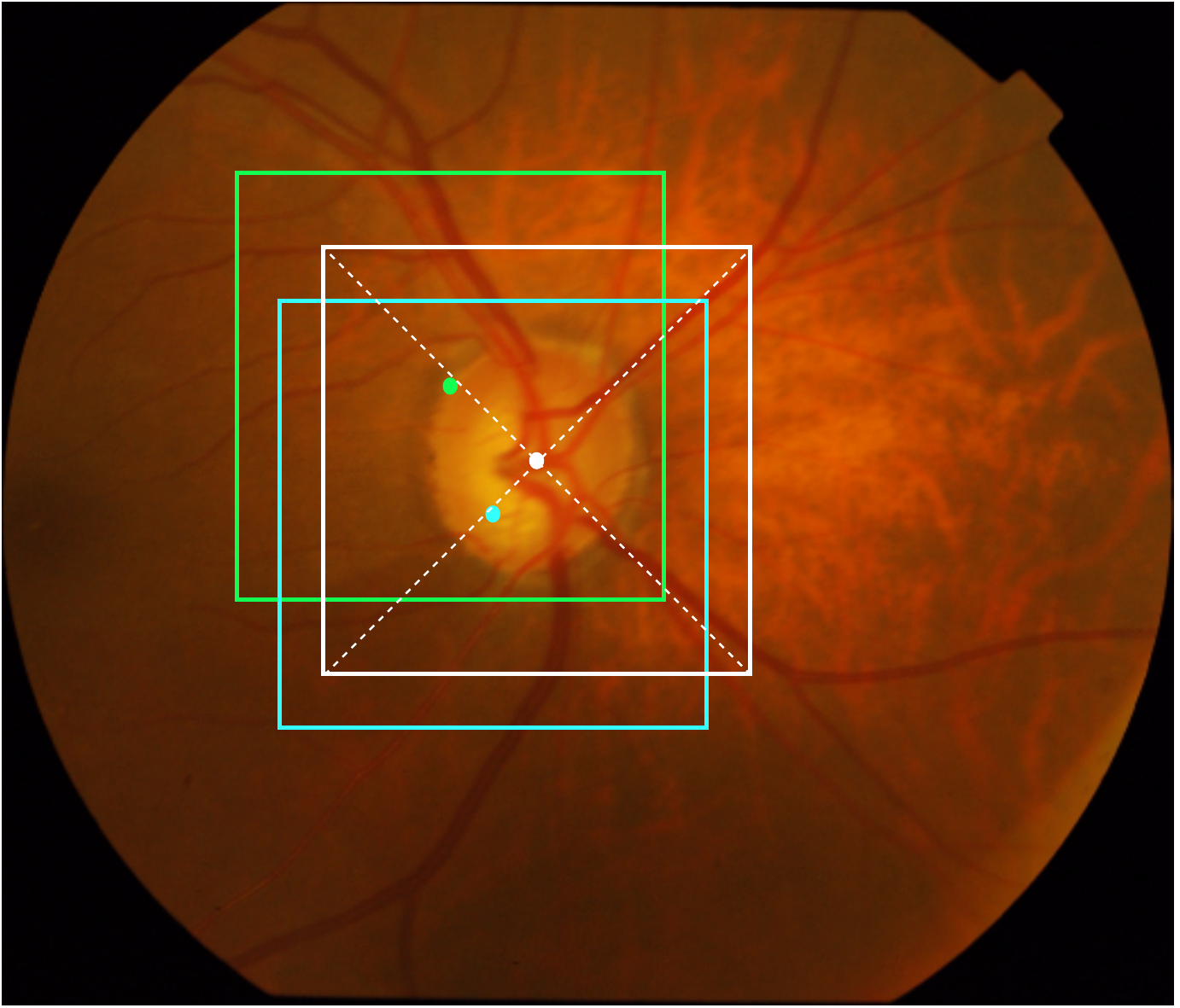}
\end{center}
\caption{\textbf{Center of mass shift to sample random patches for training phase:} We can see different patches picked at different sampling steps while training. The patch shown by white boundary has the true center of mass of the manual segmentation. Other color-coded patches are sampled by shifting the center of mass by $(a, b)$. The center of mass for each patch is color-coded.}
\label{img:center_of_mass_shift}
\end{figure}




\subsubsection{Patch selection in training}
During training, we select $P \times r$ patches encompassing the entire optic disk; we randomly vary each patch's center location. We discard patches that have less than $T$ ($T$=500 for our experiments) positive pixels in the ground truth because a network will learn little if there are very few positive pixels in a given patch. In addition to the OD, we empirically determined that the network also needs some patches belonging to the corners of the image to be more robust against borderline artifacts. In more detail, we first compute the \textit{center-of-mass} $(c_x, c_y)$ of the OD in the ground truth. Then, we randomly shift this center of mass to $(c_x + a, c_y + b)$ and take a patch of size $w \times h$. Every time we sample this new shifted center of mass, we ensure that the entire optic disk is within the original $w \times h$ patch (see Fig.~\ref{img:center_of_mass_shift}). These random shifts increase variance in the training set as each time we sample a new patch, a new region of the image is fed to the network. 




\subsubsection{Patch selection in evaluation}
There are two ways one could feed an new image to a trained model: \textit{(a).} Feed the entire image using a sliding window; \textit{(b).} Use a preprocessing technique to identify the possible locations of optic disk and just feed patches around those. Empirically, we found that feeding the entire image with a sliding window was significantly more robust and stable. As our experiments show, a trained network automatically ignores regions of the image that do not contain the OD.

\subsection{Stochastic cost functions}
\label{stochastic_loss}
Cross-entropy and dice loss are two widely used loss functions in supervised segmentation tasks. In particular, dice loss is widely used in medical image segmentation because, unlike accuracy, it is sensitive to high class imbalances (i.e., where one class is much more likely than the other). However, as detailed in \cite{10.3389/fcomp.2020.00035}, a network often perform poorly in ambiguous regions (i.e. regions with few positive pixels) when trained with these standard loss functions. Concretely, standard training leads to the network labeling only high-confidence pixels as positive. Thus, \cite{10.3389/fcomp.2020.00035} introduced a stochastic training scheme that forces the network to be more robust by randomly varying the misclassification penalty (i.e., the relative cost of a false positive vs. a false negative). In more detail, stochastic cross-entropy is defined as follows:
\begin{equation}
	\label{eq:dynamic_cross_entrpy}
	H = -\sum_{x}w{_{rand(1, \alpha, s)}}p(x)\log{q(x)},
\end{equation}
where $rand$ function draws the penalty parameter randomly from range $1$ - $\alpha$ with a step-size of $s$ to prevent exploding gradients. The stochastic version of dice loss uses the same additional parameters:
\begin{equation}
	\label{eq:dyn_dice_loss}
	F_{\beta} = (1 + B_{rand}(1, \alpha, s)^{2}) \cdot \frac{precision \cdot recall}{B_{rand}(1, \alpha, s)^{2} \cdot precision + recall}.
\end{equation}
This stochastic approach, however, has only been applied to vessel segmentation. In this paper, we applied it to OD segmentation, and, as our experiments show, verified that it is also useful for this problem.


\section{Experiments and Results}
\label{experiments_and_result}
We empirically validated our DCPA and stochastic training schemes on five retinal datasets: DRISTI \cite{dristi_paper_dataset}, DRIVE \cite{drive_dataset}, DRIONS-DB \cite{drions_db_paper_dataset}, AV-WIDE \cite{wide_dataset:6987362}, and CHASE-DB \cite{chasedb_dataset} (see Tbl.~\ref{tab:datasets}). Below, we describe our equipment and experiments in more detail.

\noindent\textbf{Hardware:} The  experiments of this paper were conducted on a Microsoft Azure VM on an Intel server with 16 cores, 512 GB RAM and 4 Tesla V100 16 GB GPUs.

\begin{table}[t!]
\caption{Dataset details}
\label{tab:datasets}
\begin{center}
\begin{small}
    \begin{tabular}{lcccccc}
        \toprule
        \rule[-1ex]{0pt}{3.5ex} \textbf{Dataset} & \textbf{ Total \#} & \textbf{Img dim.} & \textbf{OD dim.} & \textbf{$\frac{OD}{Image}$} & \textbf{$(w \times h)$} & \textbf{$(w + p) \times (h + q)$}\\

		\midrule 
        \rule[-1ex]{0pt}{3.5ex} DRISTI \cite{dristi_paper_dataset} & 101 & $2049 \times 175$1 & $380 \times 380$ & 3.13\% & $836 \times 836$ & $1040 \times 1040$\\

        \rule[-1ex]{0pt}{3.5ex} DRIVE \cite{drive_dataset} & 40 & $565 \times 584$ & $80 \times 85$ & 1.79\% & $388 \times 388$ & $572 \times 572$\\

        \rule[-1ex]{0pt}{3.5ex} AV-WIDE \cite{wide_dataset:6987362} & 30 & $1237 \times 809$ & $75 \times 80$ & 0.40\% & $388 \times 388$ & $572 \times 572$\\

        \rule[-1ex]{0pt}{3.5ex} DRIONS-DB \cite{drions_db_paper_dataset} & 110 & $600 \times 400$ & $90 \times 90$ & 3.07\% &$388 \times 388$ & $572 \times 572$\\

        \rule[-1ex]{0pt}{3.5ex} CHASE-DB \cite{chasedb_dataset} & 28  & $999 \times 960$ & $190 \times 190$ & 8.91\% & $388 \times 388$ & $572 \times 572$\\
        \bottomrule
    \end{tabular}
\end{small}
\end{center}
\end{table}

\noindent \textbf{Datasets and ground truth preparation:} As listed in Tbl.~\ref{tab:datasets}, we used five popular retinal datasets to assess our technique. The ground-truth OD segmentations for the DRISTI and DRIONS-DB datasets were available from the original authors since these two datasets were specifically designed for optic disk segmentation \cite{dristi_paper_dataset, drions_db_paper_dataset}. The optic disc is centered in these images and covers a large portion of the image. In addition, we estimated the ground-truth OD segmentations for three additional retinal datasets: DRIVE \cite{drive_dataset}, AV-WIDE \cite{wide_dataset:6987362}, and CHASE-DB \cite{ chasedb_dataset}. These datasets were originally created for vessel segmentation, so the OD ground truth was not available. Unlike the datasets created specifically for OD segmentation, the field of view in these images is not focused to the optic disk; it can appear on any corner of the image, making it a much harder problem. 




\noindent\textbf{Network training:} We used 5-fold cross validation for datasets that did not have separate training and test sets. For the ones with such separation, we used 30\% of the images in the training set for validation. We used the ADAM optimizer \cite{kingma2014method_AdamOptima}, a learning rate of $0.001$ and a mini-batch size of 8 for all experiments. We used random vertical, horizontal flips while training, and used the stochastic version of dice loss as explained in Sec.~\ref{stochastic_loss}\footnote{We also tested using cross-entropy, but the dice loss consistently yielded better results.}. 




\noindent\textbf{Network configurations:} To better understand the impact of DCPA and stochastic weights, we tested four different network/training configurations on each dataset: (1) DCPA + stochastic weights; (2) DCPA only; (3)Stochastic weights only; (4) No DCPA or stochastic weights (standard training). Additionally, we used two network sizes for each setting: $R1$, the full U-Net in the original paper \cite{Ronneberger2015UNetCN}, and $R2$, a network reduced by a factor of 2 on its width (i.e., number of channels reduced by 2 in each layer). This reduces the total parameters from ~31 million to ~7 million, making it possible to train without high end configurations. Specifically, we were able to only train the $R2$ experiments on a Intel server with two 1080 Ti GPUs (11 GB of VRAM each) but not the $R1$. In our discussion below, we refer to fixed $\beta$ vs. random $\beta$ to indicate whether a configuration used stochastic weights.


\begin{figure}[t!]
\begin{center}
    \includegraphics[width=0.7\textwidth]{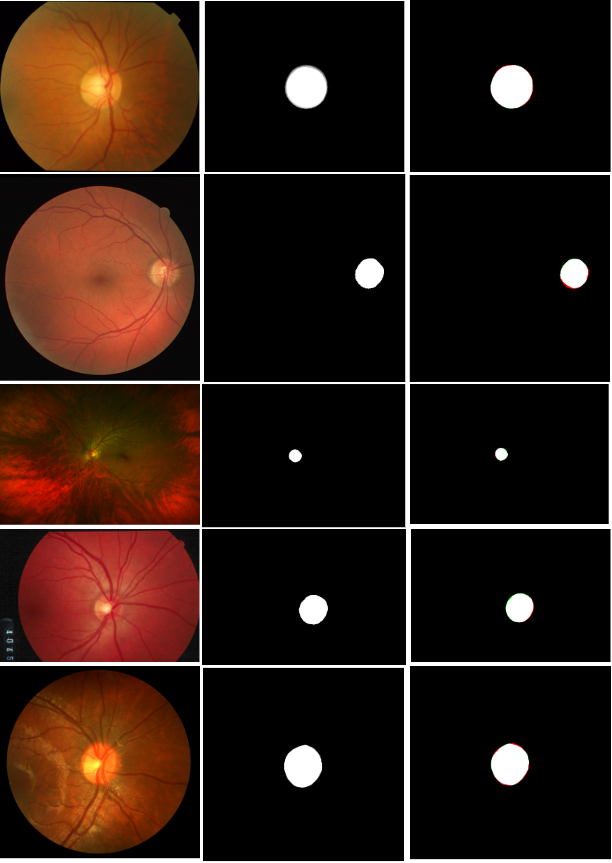}
\end{center}
\caption{\textbf{Segmentation results on different datasets(top to bottom) with image, segmentation ground truth, and our segmentation result (left to right):} DRISTI\cite{dristi_paper_dataset}, DRIVE\cite{drive_dataset}, AV-WIDE\cite{wide_dataset:6987362}, DRIONS\cite{drions_db_paper_dataset}, CHASE-DB\cite{chasedb_dataset}. The white pixels in segmentation result represent True Positive(TP)/Optic Disc, green pixels represent False Positive(FP), red pixels represent False Negative(FN), and black pixels represent True Negetive(TN)/background}
\label{img:results}
\end{figure}

\noindent\textbf{Results:} Figure~\ref{img:results} shows a sample result from each dataset, and the precision, recall, F1 score, and overlap scores for the test images are listed in Tbl.~\ref{tab:results}; the latter is a common measure for OD segmentation corresponding to $\frac{TP}{TP + FN + FP}$.  As noted above, the fixed-$\beta$ + no DCPA configuration corresponds to a regularly trained U-net; the other configurations use DCPA, stochastic weights, or both. In addition, when available, we list prior state-of-the-art results.

For DRISTI, all eight network configurations achieved excellent results. The version with the full R1 model, fixed-$\beta$ with DCPA achieved state-of-the-art results across all metrics: a precision of  0.9565, recall of 0.9505, F1 score of 0.9535, and overlap of 0.9111; the last two scores are similar to the current state of the art \cite{8451543, 6091537}. Using DCPA improved the the F1 score over $3\%$ from 0.9195 to 0.9535 and overlap score over $6\%$  from 0.851 to 0.9111 on the full R1 model. We also observed over $2\%$ improvement in F1 and overlap compared to the results reported in \cite{8451543} (F1 score of 0.897 and overlap score of 0.93)\footnote{Based on the overlap formula, the overlap should always be be lower than the F1 score, so we are not sure why overlap is larger in this study.}. Finally, in \cite{6091537} \author{Cheng et al.} report resizing their images, so we cannot directly compare our results to theirs. 

For DRIVE dataset, our DCPA-only version achieved over a $3\%$ improvement in overlap and in F1 score over the state of the art. The full R1 model had actually slightly lower values (although still above the state of the art) which means that we don't need a full network R1 and can get the similar result with smaller network R2. In either cases, we can see that DCPA plays a crucial role in improving results. Current state-of-the-art techniques \cite{7225107, Abdullah2018} reported a overlap of $0.8$ and $0.81$ respectively, whereas our technique obtained a score of $0.84$. As noted earlier, we believe that the use of DCPA enabled the network to attend more on specific region of an image where the optic disk is rather than try to encode the entire image. We can also see that using a random $\beta$ adds more robustness to the performance of a neural network|a precision of 0.89 without random $\beta$ vs 0.91 with random $\beta$, 0.86 vs 0.87 for F1 score, and 0.81 vs 0.84 for overlap.

\begin{table}[ht!]
\caption{Experimental results: The method with bold face are different ablation studies. Column \textbf{R1} referes to the full size model with ~31M parameters, whereas \textbf{R2} refers to reduced model(~7M parameters). DCPA refers to the use of Disk Centered Patch Augmentation, random beta refers to stochastic class weighted Dice loss function, and fixed beta refers to unweighted dice loss function.}
\label{tab:results}
\setlength{\tabcolsep}{1.5pt}
\begin{center}
\begin{small}
    \begin{tabular}{llcccccccc}
        \toprule 
        \textbf{Dataset} & \textbf{Method} & \multicolumn{4}{c}{R1 (31 million params.)} & \multicolumn{4}{c}{R2 (7 million params.)} \\
        \midrule
        
         & & \textbf{Precision} & \textbf{Recall} & \textbf{F$_1$} & \textbf{Overlap} & \textbf{Precision} & \textbf{Recall} & \textbf{F$_1$} & \textbf{Overlap}  \\
        \midrule

        
        \rule[-1ex]{0pt}{3.5ex}\textbf{DRISTI}&Cheng \textit{et al.} \cite{6091537} & - & - & 0.897 & 0.93 & - & - & - & -\\
        
        \rule[-1ex]{0pt}{3.5ex}&FINE-Net \cite{8451543} & - & - & \textbf{0.964} & \textbf{0.931} & - & - & - & - \\

        \rule[-1ex]{0pt}{3.5ex}&\textbf{Fixed$_\beta$ + No DCPA} & 0.9291 & 0.9101 & 0.9195 & 0.851 & 0.9468 & 0.9269 & 0.9368 & 0.8811 \\
        
        \rule[-1ex]{0pt}{3.5ex}&\textbf{Fixed$_\beta$ + DCPA} & \textbf{0.9565} & \textbf{0.9505} & \textbf{0.9535} & \textbf{0.9111} & \textbf{0.9531} & 0.9471 & \textbf{0.9501} & \textbf{0.9049} \\
        
        \rule[-1ex]{0pt}{3.5ex}&\textbf{Random$_\beta$ + No DCPA} & 0.9184 & 0.9387 & 0.9284 & 0.8664 & 0.9502 & 0.9265 & 0.9382 & 0.8836 \\
        
        \rule[-1ex]{0pt}{3.5ex}&\textbf{Random$_\beta$ + DCPA} & 0.9553 & 0.9501 & 0.9527 & 0.9096 & 0.9452 & \textbf{0.9504} & 0.9478 & 0.9008 \\
        
        \midrule
        
        \rule[-1ex]{0pt}{3.5ex} \textbf{DRIVE}&S. Roychowdhury \cite{7225107} & - & - & - & 0.8067 & - & - & - & -\\

        \rule[-1ex]{0pt}{3.5ex}&Bat Meta-heuristic \cite{Abdullah2018} & - & - & 0.8810 & 0.8102 & - & - & - & - \\
       
        \rule[-1ex]{0pt}{3.5ex}&\textbf{Fixed$_\beta$ + No DCPA} & 0.8239 & 0.9306 & 0.874 & 0.7762 & 0.8669 & 0.9266 & 0.8958 &	0.8113 \\
        
        \rule[-1ex]{0pt}{3.5ex}&\textbf{Fixed$_\beta$ + DCPA} & \textbf{0.8791} & 0.9265 & \textbf{0.9022} & \textbf{0.8218} & 0.8767 & 0.9487 &	0.9113 & 0.8371\\
        
        \rule[-1ex]{0pt}{3.5ex}&\textbf{Random$_\beta$ + No DCPA} & 0.8331 & 0.9515 & 0.8884 & 0.7992 & 0.87752 & \textbf{0.9538} & 0.9141 & 0.8418\\
        
        \rule[-1ex]{0pt}{3.5ex}&\textbf{Random$_\beta$ + DCPA} & 0.8265	& \textbf{0.9656} & 0.8906 & 0.8028 & \textbf{0.8962} & 0.9362 & \textbf{0.9158} & \textbf{0.8447} \\
        
        \midrule
        
        
        \rule[-1ex]{0pt}{3.5ex}\textbf{DRIONS-DB}&FINE-Net \cite{8451543} & - & - & 0.955 & 0.914 & - & - & - & - \\

        \rule[-1ex]{0pt}{3.5ex}&Mannis \textit{et al.} \cite{Man+16} & - & - & \textbf{0.971} & \textbf{0.944 } & - & - & - & -\\

        \rule[-1ex]{0pt}{3.5ex}&\textbf{Fixed$_\beta$ + No DCPA} & 0.9465 & \textbf{0.9494} & 0.9479 &	0.901 & 0.9637	& 0.9378 & 0.9506 & 0.9059 \\
        
        \rule[-1ex]{0pt}{3.5ex}&\textbf{Fixed$_\beta$ + DCPA} & 0.9627 & 0.9461 &	\textbf{0.9543} & \textbf{0.9126} & 0.9274 & \textbf{0.9467} &	0.937 &	0.8815 \\
        
        \rule[-1ex]{0pt}{3.5ex}&\textbf{Random$_\beta$ + No DCPA} & \textbf{0.9692} & 0.9333 & 0.9509 & 0.9064 & 0.9441 & 0.9461 & 0.9451 & 0.8959 \\
        
        \rule[-1ex]{0pt}{3.5ex}&\textbf{Random$_\beta$ + DCPA} & 0.9631	& 0.9331 & 0.9478 & 0.9008 & \textbf{0.9677} & 0.9397 & \textbf{0.9535} & \textbf{0.9111} \\
        \midrule
        
        
      \rule[-1ex]{0pt}{3.5ex}\textbf{AV-WIDE}&\textbf{Fixed$_\beta$ + No DCPA} & 0.8399 & 0.6959 & 0.7611 & 0.6143 & 0.8576 & \textbf{0.7444} & 0.797 & 0.6625 \\
       
        \rule[-1ex]{0pt}{3.5ex}&\textbf{Fixed$_\beta$ + DCPA} & 0.9115 & \textbf{0.7661} & \textbf{0.8325} & \textbf{0.7131} & \textbf{0.9287} & 0.7315 & \textbf{0.8184} & \textbf{0.6926} \\
        
        \rule[-1ex]{0pt}{3.5ex}&\textbf{Random$_\beta$ + No DCPA} & 0.7479 & 0.7348 & 0.7413 & 0.5889 & 0.8581 & 0.6926 & 0.7665 & 0.621 \\
    
        \rule[-1ex]{0pt}{3.5ex}&\textbf{Random$_\beta$ + DCPA} & \textbf{0.927} & 0.7502 & 0.8293 & 0.7084 & 0.9207 & 0.723 & 0.81 & 0.6807 \\
        
        \midrule
        
 
        \rule[-1ex]{0pt}{3.5ex}\textbf{CHASE-DB}&\textbf{Fixed$_\beta$ + No DCPA} & 0.7388 & \textbf{0.9341} & 0.8251 & 0.7023 & 0.6928 & 0.9118 & 0.7874 & 0.6493 \\
        
        \rule[-1ex]{0pt}{3.5ex}&\textbf{ Fixed$_\beta$ + DCPA} & \textbf{0.7501} & 0.9298 & \textbf{0.8303} & \textbf{0.7098} & \textbf{0.7382} & 0.9301 & \textbf{0.8231} & \textbf{0.6994} \\
        
        \rule[-1ex]{0pt}{3.5ex}&\textbf{Random$_\beta$ + No DCPA} & 0.7253 & 0.9126 & 0.8082 & 0.6782 & 0.7305 & 0.9286 & 0.8177 & 0.6916  \\
        
        \rule[-1ex]{0pt}{3.5ex}&\textbf{Random$_\beta$ + DCPA} & 0.7303 & 0.9305 & 0.8183 & 0.6925 & 0.7178 & \textbf{0.9441} & 0.8156 & 0.6886 \\
        \bottomrule
    \end{tabular}
\end{small}
\end{center}
\end{table}

For DRIONS-DB \cite{drions_db_paper_dataset}, we can see that using DCPA yielded slightly better performence than FineNet \cite{8451543}, the current state of the art for this dataset. Even With the smaller R2 model, we obtained comparable results (0.9111 vs 0.9140 F1 score). In this case, using a random $\beta$ in conjunction with DCPA yielded the best results. 


For the AV-WIDE dataset \cite{wide_dataset:6987362}, the F1 score and overlap improved significantly with the use of DCPA (0.76 vs 0.83 F1, and 0.61 vs 0.71 IOU in the full model R1) (Table \ref{tab:results}). This dataset was created for vessel segmentation and classification tasks. As such, it is more complex than the other datasets and has a much smaller disk-to-image ratio. Here, the OD only covers around 6\% of the image patches as opposed to around 17\% for the other datasets, making the problem harder. For the R2 model, we also observed an improvement of $2\%$ on the F1 score when using the fixed-$\beta$ and and improvement of about $5\%$ on the F1 score when using random-$\beta$. Finally, the results on this dataset show that our technique is robust to different OD-to-image ratios, confirming that it generalizes better than other state-of-the-art techniques. 

For CHASEDB \cite{chasedb_dataset}, we can see that DCPA yielded more than $2\%$ better recall (0.7388 vs 0.7501), about $1\%$ better F1 score (0.8251 vs  0.8303), and a similar overlap (0.7023 vs 0.7098) compared to standard training (Table \ref{tab:results}). In addition, the F1 score improved $5\%$ by using DCPA on the reduced model R2.

\begin{table}[t!]
\caption{Experimental running time in seconds for $R1$ and $R2$ in the same setting. The run-time reported is for one experiment for each datasets conducted in table \ref{tab:results} for full 351 epochs with batch size 8.}
\label{tab:timeresults}
\setlength{\tabcolsep}{1.5pt}
\begin{center}
\begin{small}
    \begin{tabular}{llccccccc}
        \toprule
        \textbf{Model} & \textbf{DRISTI} & \textbf{DRIVE} & \textbf{DRIONS-DB} & \textbf{AV-WIDE} & \textbf{CHASE-DB} \\
        \midrule

        
        \rule[-1ex]{0pt}{3.5ex}\textbf{R1}& 14,200 & 1,530 & 3,760 & 21,800 & 21,600 \\
        
         \rule[-1ex]{0pt}{3.5ex}\textbf{R2}& 7,250 & 1,170 & 2,390 & 15,600 & 15,500 \\
         
        \midrule
        \bottomrule
    \end{tabular}
\end{small}
\end{center}
\end{table}

\noindent \textbf{Time results:} Table~\ref{tab:timeresults} show the running time in seconds for the R1 and R2 network configurations on the various datasets. As expected, the running time for training of the smaller models was considerably lower than the larger ones.

\section{Discussion}
\label{sec:discussion}
We have shown that by using a novel training technique, we can consistently achieve state-of-the-art results using relatively small networks. This is very important because most researchers around the world do not have access to high-end training clusters as large labs and companies do. We can see in Tbl.~\ref{tab:results} that DCPA improved or matched the results compared to a full U-Net network. We have shown such results in two of the widely used public dataset(DRISTI and DRIONS) and generated our own OD ground truth for three other datasets (DRIVE, AV-WIDE, and CHASEDB). As detailed in Sec.~\ref{sec:methodology}, DCPA restricts training to patches that contain the optic disk at different positions (to account for center-of-mass shifts). This ensures that the network is robust to OD position, as shown in our various results.

In particular, we obtained excellent results on datasets with the OD in the center (DRISTI, CHASE-DB, STARE, DRIONS), the right or left side (DRIVE), or with a very small disk-to-image ratio (AV-WIDE). 

In addition, the random-beta stochasticity described in paper \cite{10.3389/fcomp.2020.00035} enables the network to be more robust against ambiguous pixels around the border of the OD, which consistently improved recall as shown in Tbl.~\ref{tab:results}. For example, we can see that stochasticity, along with DCPA, yielded higher recall on DRIVE. However, for other datasets, such as DRIONS-DB, and CHASE-DB, the use of DCPA was sufficient without stochasticity. As mentioned earlier, in vessel segmentation \cite{10.3389/fcomp.2020.00035} the vessels are well scattered across the image. Thus, the network needs to learn a wide variation of shapes, which made stochastic penalties very useful. This robustness was not as crucial in the case of the optic disc, since it is usually a blob-like shape in a fixed location. As such, the use of DCPA was often sufficient to achieve state of the art results.
 



\section{Conclusions and Future work}
\label{sec:conclusions}
In our work, we developed a novel approach for segmenting the optic disc in retinal fundus images. Our extensive comparison with the state of the art, across five datasets, showed that our technique consistently achieves excellent precision, recall, F1, and overlap scores. In addition, our ground-truth segmentations for three of the datasets (CHASE-DB, DRIVE, AV-WIDE) can be used as a benchmark for future researchers. Overall, our results suggest that it is important to have patches that contain the entire optic disk, with variations in positions, in order for the network robustly segment the OD across different imaging devices, zoom factors, image quality, and device resolution. For future work, we aim to use this work for diagnosing glaucoma and other retinal diseases that affect the optic disc.





\section*{Acknowledgments}
This research was funded in part by NSF award 1849946. We also thank Michael Allingham at the Duke University Medical Center for validating our ground truth for the DRIVE, AV-WIDE, and CHASE-DB datasets. 





\bibliographystyle{plain} 
\bibliography{report}

\end{document}